\documentclass{egpubl}
\usepackage{sca2026}

\SpecialIssuePaper         %

\CGFccby

\usepackage[T1]{fontenc}
\usepackage{dfadobe}  

\usepackage{cite}  %
\BibtexOrBiblatex
\electronicVersion
\PrintedOrElectronic
\ifpdf \usepackage[pdftex]{graphicx} \pdfcompresslevel=9
\else \usepackage[dvips]{graphicx} \fi

\usepackage{egweblnk}

\usepackage{amsfonts}
\usepackage{amsmath}
\usepackage{pifont}
\usepackage{booktabs}
\usepackage{xcolor}
\usepackage{multirow}
\usepackage[normalem]{ulem}
\usepackage[super]{nth}
\usepackage{cleveref}
\usepackage{colortbl}
\usepackage{tabularray}
\usepackage{listings}
\usepackage[ruled,vlined]{algorithm2e}
\usepackage[labelfont=bf,textfont=it,skip=6pt]{caption}
\usepackage{subcaption}
\usepackage{xspace}

\newif\ifdraft
 \draftfalse
\ifdraft
\newcommand{\myc}[1]{{\color{blue}\textbf{Mingyi:} #1}}
\newcommand{\xlc}[1]{{\color{magenta}\textbf{Xuelin:} #1}}
\newcommand{\tkc}[1]{{\color{cyan}\textbf{Taku:} #1}}
\newcommand{\ruic}[1]{{\color{green}\textbf{Rui:} #1}}

\else
\newcommand{\myc}[1]{}
\newcommand{\xlc}[1]{}
\newcommand{\tkc}[1]{}
\newcommand{\ruic}[1]{}

\fi

\newif\ifhighlight
\highlightfalse
\ifhighlight
\newcommand{\rev}[1]{{\color{blue}#1}}
\else
\newcommand{\rev}[1]{#1}
\fi

\newcommand{\keep}[1]{}
\newcommand{\old}[1]{}

\providecommand{\citet}[1]{\cite{#1}}
\providecommand{\citeyearpar}[1]{\cite{#1}}

\title[Hand Motion Completion]%
      {Prior-First, Condition-Second: Scalable and Controllable Hand Motion Completion}

\author[M.~Shi, X.~Chen \& T.~Komura]
{\parbox{\textwidth}{\centering
        Mingyi Shi$^{1}$\orcid{0000-0002-5180-600X}
        , Xuelin Chen$^{2}$
        and Taku Komura$^{1}$\orcid{0000-0002-2729-5860}
        }
        \\
{\parbox{\textwidth}{\centering
        $^1$The University of Hong Kong, Hong Kong\\
        $^2$Adobe Research, London, UK
       }
}
}

\begin{document}

\teaser{
  \includegraphics[width=0.9\linewidth]{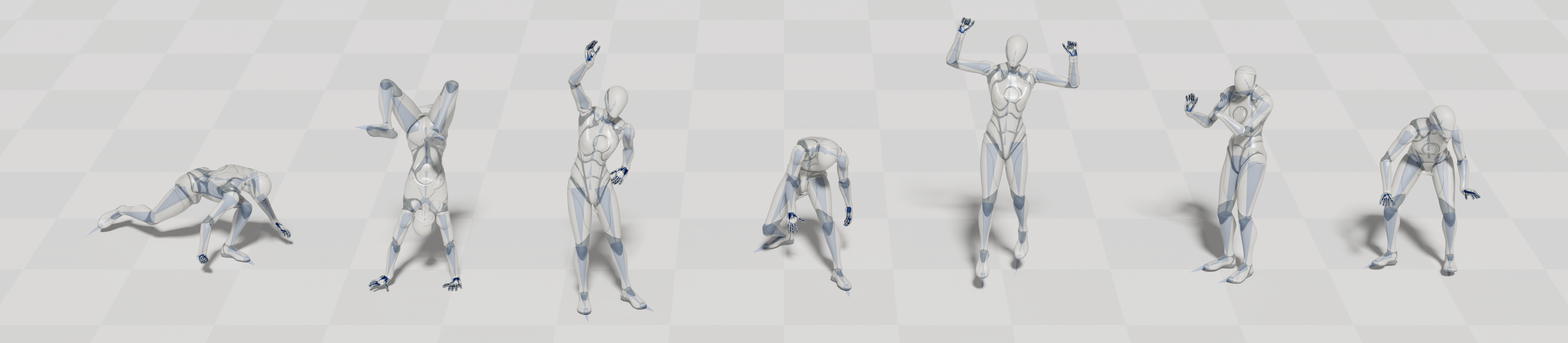}
  \centering
  \caption{Our method robustly completes high-fidelity hand motion conditioned on global body dynamics and optional semantic controls. Instead of training a fully end-to-end conditioned generator, we constrain the outputs to a physically plausible kinematic manifold, resulting in coherent articulation and strong robustness across diverse inputs.}
  \label{fig:teaser}
}

\maketitle

\begin{abstract}

Synthesizing hand motion that matches the full body motion and the semantic labels is a difficult task due to their high degrees of freedom and the lack of semantic labels. 
To cope with this issue, 
we propose a \emph{prior-first, condition-second} framework for body-conditioned hand motion completion. Our framework first learns a generic body--hand kinematic prior from large-scale unstructured and unlabeled motion data, capturing the intrinsic coordination between global body dynamics and hand articulation. Semantic control is then introduced through lightweight adaptation on top of the frozen prior, avoiding the need to relearn kinematic structure for each control interface.    
Our framework centers on a streaming, autoregressive body–hand prior that generates coherent, kinematically consistent hand motion from body dynamics in real time, using structured kinematic modeling to maintain mechanical body–hand coupling.
To enable practical controllability under limited supervision, we introduce semantically-layered adapters that inject conditioning signals at appropriate kinematic levels, supporting both self-supervised attribute control and weakly supervised text-driven control with only a few hours of labeled data. Extensive evaluations demonstrate that our framework improves kinematic plausibility, robustness, and controllability compared to end-to-end conditioned baselines, particularly in low-resource and cross-dataset settings. We further showcase real-time inference and an interactive authoring workflow, highlighting the applicability to production animation pipelines.


\end{abstract}

\section{Introduction}

Hand motion is a critical channel for conveying intent, emphasis, and affect in human behavior. Subtle variations in finger articulation can communicate nuances such as hesitation, confidence, or invitation, even  the body motion remains unchanged. In production animation pipelines, however, hands remain among the most challenging components to animate because they exhibit high degrees of freedom.
Even advanced tools such as Cascadeur~\cite{Cascadeur} can only provide single-frame AutoPosing, which still depends on extensive manual authoring or specialized capture setups.

While recent progress has been made in data-driven human motion generation, most existing methods focus on body motion and do not explicitly model hand motion, leaving high-quality and controllable hand motion synthesis an open challenge.
A central difficulty lies in the nature of the supervision available for generative hand motion modeling. 
First, while motion capture systems have become increasingly accessible, paired semantic supervision (e.g., text descriptions, audio cues, or high-level attributes of the motion) remains expensive and challenging to collect at scale.
Second, motion semantics are inherently imbalanced and long-tailed: a small number of frequent actions and poses dominate existing annotations, whereas subtle yet critical variations in timing, articulation, and intent are rarely labeled.
Moreover, motion annotations such as text and audio are highly dataset-dependent. Variations in capture protocols, skeleton definitions, annotation schemes, and even cultural context lead to semantic labels that do not transfer reliably across datasets.
While full-body motion synthesis already suffers from these annotation bottlenecks, the problem is even more critical for hands, where semantic labels are sparser and cross-dataset retargeting is notoriously difficult.

Motivated by these observations, we argue that controllable motion generation should be framed as sampling from a learned kinematic prior, rather than learning an end-to-end mapping from sparse conditions to motion trajectories. The kinematic structure of motion—how body parts coordinate, transfer momentum, and evolve over time—should be learned independently under minimal semantic assumptions using abundant unlabeled motion data, and semantic control can then be implemented as a lightweight constraint that guides the sampling process on this pre-learned manifold.

Crucially, this separation also addresses the lack of cross-dataset semantic consistency: by learning kinematic structure independently of any particular annotation scheme, the resulting prior captures motion regularities that are invariant across datasets, skeletons, and labeling conventions. 
Semantic control can then be introduced as a lightweight, dataset-specific conditioning on top of this shared prior. 
This perspective naturally leads to a \emph{prior-first, condition-second} paradigm. \rev{While prior learning and adapter-based conditioning are individually established, our contribution is their concrete instantiation for body-conditioned hand motion: a streaming autoregressive body--hand prior, a kinematic chain cascading attention enforcing body-to-hand coupling, and semantically-layered adapters that inject control at the most effective kinematic levels. }

To realize this paradigm, we present a unified framework for body-conditioned hand motion completion. At its core, our framework learns an autoregressive kinematic prior that completes temporally coherent and kinematically consistent hand motion given global body dynamics. The key design choice is to model hand motion as a conditional, streaming process: at each time step, a short future clip of hand pose is generated via a diffusion model conditioned on the current body state and a short history of past hand motion. This autoregressive formulation supports online inference and enables real-time use while maintaining long-term temporal consistency.
Within this framework, we incorporate a kinematic chain cascading attention mechanism to model body-to-hand dependencies efficiently. Instead of treating body joints as isolated tokens, this mechanism follows the articulated structure of the human body and aggregates information along the kinematic chain (Root $\rightarrow$ Spine $\rightarrow$ Limbs). This structured attention encourages generated hand motions to remain mechanically coupled to upstream body dynamics, reducing common artifacts such as phase drift or implausible wrist behavior.

To support practical high-level control without requiring massive paired supervision, we keep the kinematic prior frozen and introduce semantically-layered adapters for conditional generation. Observing that semantic commands naturally operate at different kinematic levels (e.g., global style versus local finger articulation), we inject conditioning signals into specific stages of the cascading prior. This design enables data-efficient adaptation: only a few hours of labeled data are required to train the adapters. We instantiate this framework with two control strategies: (i) self-supervised parameter control using numerically computed motion attributes, and (ii) text-driven control using structured captions generated by vision-language models. In both cases, we employ diversity-aware subset selection to maximize coverage of the learned kinematic manifold under limited supervision.

We evaluate the proposed framework across diverse control interfaces and motion sources, and demonstrate its practical utility through a real-time inference pipeline and an interactive add-on. Our main contributions include: 

\begin{itemize} 
    \item We propose a \emph{prior-first, condition-second} framework for controllable hand motion completion, in which a generic kinematic prior is learned from large-scale unstructured motion data and semantic control is introduced via lightweight adaptation.
    \item We present a streaming, autoregressive prior that enables temporally coherent and kinematically consistent hand motion completion conditioned on body dynamics.
    \item We introduce data-efficient, semantically-layered adaptation on top of prior, supporting both self-supervised parameter control and weakly supervised text control with only a few hours of labeled data.

\end{itemize}

\section{Related Work}
\label{sec:related}

\subsection{Hand Motion Synthesis and Completion}

Generating hand motion has been an active research topic across graphics and robotics. Early work often relied on physics-based simulation and carefully designed constraints to synthesize realistic hand--object manipulation \cite{pollard2005physically, liu2008synthesis, ye2012synthesis, kry2006interaction, liu2009dextrous}. A notable data-driven alternative is the motion-graph approach of J\"{o}rg et al.~\cite{jorg2012data}, which retrieves and stitches finger motion clips from a database using body features as a query. \rev{Earlier, Majkowska et al.~\cite{majkowska2006automatic} similarly spliced separately captured hand animations onto body motion.} While effective for small, constrained domains such as co-speech gestures, retrieval-based methods face fundamental scalability limitations: memory cost grows linearly with the database, graph optimization for temporal continuity is slow, and stitching introduces visible artifacts at segment boundaries under large body dynamics. With the increasing availability of hand motion datasets \cite{taheri2020grab, wang2024furelise, liu2022beat, zhang2024both2hands} and the rapid progress of generative models, learning-based approaches that model a continuous distribution over hand motions have become increasingly scalable and popular. Existing learning-based methods are typically trained end-to-end on scenario-specific datasets and differ mainly in their target applications and conditioning signals. A major branch focuses on \textit{hand-only} generation, emphasizing dexterous finger articulation for HOI \cite{zhang2021manipnet, zhou2022toch, zheng2023cams, christen2024diffh2o} or music performance \cite{wang2024furelise, wu2023marker, gan2024pianomotion10m}; these methods often provide high-fidelity hand motion but lack full-body context for body--hand coordination. Another line treats the body as a strong prior and formulates hand motion as a completion problem. Body2Hands \cite{ng2021body2hands} predicts 3D hand motion over time from upper-body motion in conversational settings, but does not address arbitrary full-body motions nor semantic control. More recently, BOTH2Hands \cite{zhang2024both2hands} captures and annotates a paired full-body+hand dataset (about 8 hours) and trains a diffusion model to generate hand motion conditioned on body dynamics and text.

A critical challenge across these lines is that datasets are typically collected for a specific scenario with specialized setups and are used in isolation. Meanwhile, high-quality full-body mocap resources are rapidly growing, covering locomotion~\cite{shi2025motionpersona}, dyadic interactions~\cite{ho2025interact}, and co-speech gestures \cite{liu2022beat}. 
Even when multiple body+hand datasets exist, as shown in Table~\ref{tab:dataset_landscape}, their annotations are rarely interoperable across domains, making it difficult to reuse supervision to train a single controllable model across scenarios.
This fragmentation encourages methods that are tightly coupled to a particular capture protocol or interaction context, and limits the scalability of end-to-end text-to-hand training; for example, BOTH2Hands  \cite{zhang2024both2hands} still re-collects and re-annotates paired body--hand motion and trains its text-conditioned model on only $\sim$8 hours of labeled data. These observations motivate our framework: instead of attempting to standardize heterogeneous supervision, we unify kinematic learning under a shared body--hand prior trained from large-scale unlabeled motion, and introduce task-specific control via lightweight adapters.

\rev{Hand motion is also studied under related formulations. Some works forecast hand motion from body or egocentric context~\cite{qi2023diverse, tang2024prompting, hatano2025invisible}, while others model the body and hands jointly~\cite{ding2024expressive, duran2026fusion, taheri2022goal}. In contrast to these forecasting and joint-generation settings, we target body-conditioned hand completion and separate a reusable kinematic prior from lightweight semantic adapters, letting us exploit large-scale unlabeled body motion and add control with little paired data.}

\begin{table}[b]
    \centering
    \caption{\textbf{Fragmented supervision.} Hand-related datasets are collected for distinct tasks with non-interoperable annotations (speech vs.\ music/MIDI; hand-centric vs.\ action/object text), limiting cross-dataset reuse for conditional hand motion generation.}
    \resizebox{\columnwidth}{!}{
    \begin{tabular}{p{0.33\linewidth} c c p{0.5\linewidth}}
    \toprule
    \textbf{Dataset / Work} & \textbf{Hours} & \textbf{Include body} & \textbf{Annotation modality} \\
    \midrule
    FürElise~\citeyearpar{wang2024furelise} & 10 & \texttimes & audio, Piano MIDI \\ \addlinespace
    GRAB~\citeyearpar{taheri2020grab} & 3.75 & \checkmark & action text, objects \\ \addlinespace
    BEAT~\citeyearpar{liu2022beat} & 76 & \checkmark & speech \\ \addlinespace
    InterAct~\citeyearpar{ho2025interact} & 10 & \checkmark & action, speech \\ \addlinespace
    MotionPersona~\citeyearpar{shi2025motionpersona} & 50 & \checkmark & action, body shape \\ \addlinespace
    BOTH2Hands~\citeyearpar{zhang2024both2hands} & 8.31 & \checkmark & text description \\
    \bottomrule
    \end{tabular}}
    \label{tab:dataset_landscape}
\end{table}

\subsection{Motion Generative Model as Prior}

Learning a strong motion prior and reusing it across tasks has raised more interests. Existing models adopt different probabilistic formulations, including distance-field \cite{tiwari2022pose}, normalizing flows \cite{Henter_2020}, and VAEs \cite{ling2020character}. For temporally structured priors, HuMoR \cite{rempe2021humor} proposes an autoregressive CVAE over pose dynamics and demonstrates strong results for pose estimation, though frame-level autoregression can suffer from error accumulation and limited long-horizon coherence. More recently, diffusion models have shown strong capability in capturing high-fidelity, multimodal motion distributions, and autoregressive diffusion priors such as CAMDM \cite{camdm} and AMDM \cite{amdm} enabling streaming inference and real-time character control.

In parallel, several works study how to steer or adapt a pretrained motion prior with lightweight conditioning, reducing the need for large paired datasets. PriorMDM \cite{shafir2024human} explores finetuning diffusion-based motion models for two-person generation and spatio-temporal control; attention-based conditioning has also been used for motion editing and style transfer \cite{raab2024monkey}, and parameter-efficient updates (e.g., LoRA) can improve adaptation efficiency under limited supervision \cite{sawdayee2025dance}. 
Our work builds on this direction and instantiates it for body-conditioned hand motion completion: we learn a reusable body--hand kinematic prior from large-scale unlabeled motion, and introduce controllable generation via semantically-layered adapters, enabling cross-modal control with limited paired supervision. 

\section{Method}
\label{sec:method}

We aim to synthesize high-fidelity hand motions $\mathbf{x} \in \mathbb{R}^{T \times D_h}$ that remain kinematically consistent with the global body dynamics $\mathbf{b} \in \mathbb{R}^{T \times D_b}$, where $T$ is the number of frames, $D_h$ and $D_b$ are the dimensions of the hand and body motion, respectively. 
Rather than learning a direct mapping from sparse conditions to detailed hand motion, which can be data-hungry and prone to physically inconsistent results, we propose a two-stage framework:
(1) pre-training a robust autoregressive body-hand prior that learns the conditional kinematic distribution $p(\mathbf{x}\mid\mathbf{b})$ via a physically motivated cascading architecture; and
(2) learning a lightweight conditional adapter that adds optional, prompt-driven control on top of the frozen prior using limited labeled data.
This design makes the body-conditioned prior the core modeling component, and uses downstream conditioning tasks as a practical way to evaluate and utilize the prior without requiring massive paired datasets.

\subsection{Kinematic Representation}
\label{subsec:representation}

A key design choice in our framework is the explicit decoupling of body conditioning signals from hand motion targets. 
From the full-body motion represented by a joint chain, we extract the kinematic state $\mathbf{b}_n$ and the hand state $\mathbf{x}_n$ for each frame $n$.

\paragraph{Body state input}
For the global body motion $\mathbf{b}$, we utilize a hybrid representation that captures both spatial occupancy and skeletal articulation. Specifically, for frame $n$, we extract a feature vector $\mathbf{b}_n$ consisting of: joint positions $\mathbf{p}_n \in \mathbb{R}^{J_b \times 3}$,  velocities $\mathbf{v}_n \in \mathbb{R}^{J_b \times 3}$, and local rotations $\mathbf{q}_n \in \mathbb{R}^{J_b \times 6}$ (represented as 6D continuous representations~\cite{Zhou_2019_CVPR}). It includes hand joint positions but excludes hand local rotations, which are the prediction target.
To ensure numerical stability, positions and velocities are normalized by the dataset mean and standard deviation, while local rotations are kept in their raw format to preserve geometric properties.

\paragraph{Hand State Target.}
The hand state $\mathbf{x}_n$ concatenates the 6D rotations of both hands.
We treat the wrist (hand root) differently from the finger joints: fingers are predicted in their local coordinate frames, while the wrist is predicted in global orientation so that its prediction is decoupled from the arm's varying orientation.
To reintegrate the hand into the full-body hierarchy during inference, the predicted global wrist orientation is converted back to the local frame of its parent joint (the forearm) via:

\begin{equation}
    q_{wrist}^{local} = (q_{parent}^{global})^{-1} \cdot q_{wrist}^{global}
\end{equation}

This ensures the synthesized hand remains kinematically consistent with the body skeleton, regardless of the arm's orientation.

\begin{figure*}[!ht]
    \centering
    \includegraphics[width=\linewidth]{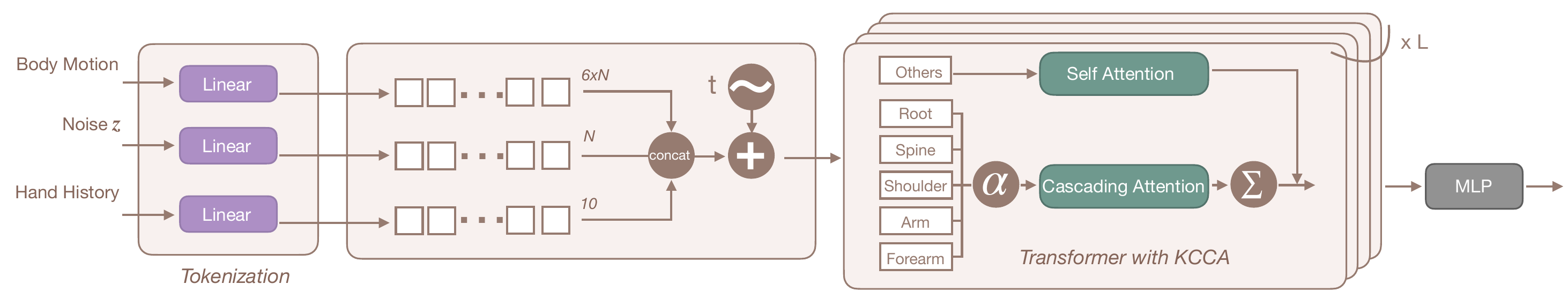}
    \caption{Overview of our diffusion block in the autoregressive prior. We project the body motion window and a 10-frame history buffer ($P{=}10$) into condition tokens, and the noisy hand motion into latent tokens. The sequence is passed through the Transformer backbone to generate an output window of $N$ frames (the prediction window of length $L$ in the text). Tokens along the root-to-wrist chain are aggregated via kinematic chain cascading attention (KCCA) to form the kinematic context.}
    \label{fig:network}
\end{figure*}

\subsection{Autoregressive Prior with Transformer-based Diffusion}
\label{subsec:prior}

To balance the temporal coherence while maintaining high-fidelity and diverse motion generation, we adopt a clip-level autoregressive formulation.
Each update step of the model predicts a fixed-length window of hand motion, and the model is trained to predict the future motion of the window given the current body motion and the past hand motion. 

\paragraph{Formulation.}
Let $L$ denote the prediction window size (shown as $N$ in Figure~\ref{fig:network}) and $P$ denote the length of the conditioning history. At any time step $t$, we aim to synthesize a hand motion sequence $\mathbf{x}_{seq} \in \mathbb{R}^{L \times D_h}$ that covers the current and future frames $[t, t+L]$.
The generation is conditioned on:
(1) The corresponding body motion window $\mathbf{b}_{seq} \in \mathbb{R}^{L \times D_b}$, which provides the full-body kinematic context; and
(2) The history buffer of past hand motions $\mathbf{h}_{past} \in \mathbb{R}^{P \times D_h}$, covering frames $[t-P, t]$.
This formulation allows the model to generate arbitrary-length future hand motion while ensuring continuity with historical motion. Temporal causality is enforced at the system level: each prediction window conditions only on the current body motion and the $P$-frame hand history from the previous output; attention is bidirectional within the current window, but no future frames from subsequent windows are accessible. To achieve the desired prediction, we adopt a Transformer-based diffusion model $\epsilon_\theta$ to predict the hand motion sequence from a noise $\boldsymbol{z} \in \mathbb{R}^{L \times D_h}$:

\begin{equation}
    \mathbf{x}_{seq} = \epsilon_\theta(\boldsymbol{z}, \mathbf{b}_{seq}, \mathbf{h}_{past})
\end{equation}

\paragraph{Transformer-based Diffusion.}
As shown in Figure~\ref{fig:network}, in the Transformer architecture, all heterogeneous data streams are treated as sequences of tokens, processed via attention layers to generate an output sequence of equal length.
Therefore, our network first projects all inputs into a unified latent space. 
We employ MLP encoders, $\mathcal{E}_{body}$ and $\mathcal{E}_{hist}$, to map the body motion window $\mathbf{b}_{seq}$ and history buffer $\mathbf{h}_{past}$ into condition tokens. Similarly, the noisy hand motion $\mathbf{x}^{(\tau)}_{seq}$ is linearly projected into latent tokens.
These tokens are concatenated to form the input sequence $\mathbf{Z}_{in}$, augmented with sinusoidal positional embeddings ($\mathbf{PE}$) to preserve temporal order:
\begin{equation}
    \mathbf{Z}_{in} = \text{Concat}\Big( \mathcal{E}_{body}(\mathbf {b}_{seq}), \mathcal{E}_{hist}(\mathbf{h}_{past}), \text{Linear}(\mathbf{x}^{(\tau)}_{seq}) \Big) + \mathbf{PE}
\end{equation}
The body motion is projected into 6 tokens per frame: one frame-level full-body token and 5 kinematic tokens from the root-to-wrist chain.
The sequence $\mathbf{Z}_{in}$ is then passed through the Transformer. To handle the spatio-temporal complexity, we interleave two attention mechanisms: 
(1) Temporal self-attention, which models dependencies along the sequence length (time axis); and 
(2) Kinematic hierarchical attention (Sec.~\ref{subsec:attention}), which models the spatial coupling between hand tokens and body tokens along the kinematic chain.

Finally, a linear decoder projects the output tokens back to the motion dimension. Following recent best practices in motion synthesis~\cite{camdm, mdm}, we predict the clean signal $\hat{\mathbf{x}}_0$ directly rather than the noise $\epsilon$ or velocity. This signal prediction objective facilitates better geometric consistency.

\subsection{Kinematic Chain Cascading Attention}
\label{subsec:attention}

Previous works typically process the full body as a flattened sequence of all joints, treating the relationship between the hands and the feet as equivalent to that between the hands and the elbows. However, biological motion is driven by kinetic energy transfer along a specific kinematic chain. To explicitly model this mechanical coupling, we propose Kinematic Chain Cascading Attention (KCCA), an attention mechanism that operates along the root-to-wrist chain to model this energy transfer.

\paragraph{Dynamic Gating.}
Considering the ordered set of kinematic tokens
$\mathcal{T}_{chain} = \{ \mathbf{c}_{root}, \mathbf{c}_{spine}, \mathbf{c}_{shoulder}, \mathbf{c}_{arm}, \mathbf{c}_{forearm} \}$, 
we structure the synthesis process to mirror the biological hierarchy. 
Rather than attending to all body parts indiscriminately, the hand query $\mathbf{Q}_n$ (derived from the noisy hand state at frame $n$) dynamically distributes its attention along the kinematic chain.
Formally, we calculate a context-aware importance distribution $\boldsymbol{\alpha}$ via a lightweight gating network:
\begin{equation}
    \boldsymbol{\alpha} = \text{Softmax}\left( \text{MLP}_{gate}([\mathbf{Q}_n ; \mathbf{c}_{root}]) \right).
\end{equation}
Here, the gating network conditions on both the local hand query and the global root momentum to decide which kinematic level is dominant for the current frame. 
Crucially, in the early stages of reverse diffusion where $\mathbf{Q}_n$ is dominated by noise, the dependence on the clean global body state $\mathbf{c}_{root}$ allows our MLP gate to maintain stable kinematic focus, unlike standard attention mechanisms which often degrade into random correlations.
\rev{In other words, the key difference from standard self-attention is structural: rather than letting the noisy hand query attend freely over all body tokens (which provides no inductive bias and is especially unreliable at high noise levels), KCCA restricts attention to the ordered root-to-wrist chain and gates it with the clean root state.}

\paragraph{Hierarchical Aggregation.}
The final kinematic context $\mathbf{h}_{out}$ is aggregated via a hierarchical gated summation:
\begin{equation}
    \mathbf{h}_{out} = \sum_{\mathbf{c}_k \in \mathcal{T}_{chain}} \alpha_k \cdot \text{Attention}(\mathbf{Q}_n, \mathbf{c}_k).
\end{equation}
By constraining the attention scope to this specific chain and weighting it via $\boldsymbol{\alpha}$ (where $\alpha_k$ is the weight for token $\mathbf{c}_k$), we explicitly enforce the biomechanical constraints of energy transfer.
For instance, during high-dynamic motions (e.g., jumping), the gate learns to upweight upstream tokens ($\mathbf{c}_{root}, \mathbf{c}_{spine}$) to capture global momentum; conversely, for fine-grained interactions, it shifts focus downstream to $\mathbf{c}_{forearm}$. Without such explicit momentum coupling, small local rotation errors can accumulate along the kinematic chain, causing hands to visually ``float'' or slide relative to the body even when the skeleton hierarchy is correct. KCCA mitigates this by anchoring the generation to upstream dynamics at every diffusion step.

\subsection{Training and Inference}
\label{subsec:training}

\paragraph{Training Objectives.}
We train the diffusion transformer by minimizing a composite loss:
\begin{equation}
    \mathcal{L}_{total} = \mathcal{L}_{rec} + \lambda_{geo} \cdot \mathcal{L}_{geo} + \lambda_{sym} \cdot  \mathcal{L}_{sym}.
\end{equation}

\noindent\textit{Rotation and velocity reconstruction.}
We supervise both per-frame joint rotations and their finite-difference velocities. Defining $\Delta \mathbf{x}_{n}=\mathbf{x}_{n}-\mathbf{x}_{n-1}$ (and similarly $\Delta \hat{\mathbf{x}}_{0,n}$):
\begin{equation}
    \mathcal{L}_{rec}
    =
    \frac{1}{L}\sum_{n=1}^{L}\left(
    \left\|\hat{\mathbf{x}}_{0,n}-\mathbf{x}_{n}\right\|_{1}
    +
    \left\|\Delta \hat{\mathbf{x}}_{0,n}-\Delta \mathbf{x}_{n}\right\|_{1}
    \right).
\end{equation}

\noindent\textit{FK-based geometric consistency.}
Let $\mathrm{FK}(\cdot;\mathcal{S})$ denote the forward-kinematics operator for skeleton $\mathcal{S}$. Given the body state $\mathbf{b}_n$ and hand rotation $\mathbf{x}_n$, FK converts local rotations to global 3D joint positions $\mathbf{y}_{n}\in\mathbb{R}^{J_h\times 3}$. We penalize the position error:
\begin{equation}
    \mathcal{L}_{geo}
    =
    \frac{1}{LJ_h}\sum_{n=1}^{L}\sum_{j=1}^{J_h}
    \left\|\hat{\mathbf{y}}_{n,j}-\mathbf{y}_{n,j}\right\|_{2}^{2}.
\end{equation}

\noindent\textit{Dual-hand relative distance loss.}
To ensure correct spatial coordination between the two hands, which is needed for motions such as clapping hands,  
we define a loss that evaluates the relative distance between corresponding left/right end-effectors
with respect those of the ground truth.
. \rev{Here $\mathcal{J}_{dual}$ denotes a \emph{set} of paired left/right end-effector joints (i.e., the five fingertips together with the wrist), so the loss matches several inter-hand distances simultaneously rather than a single global hand-to-hand distance.} For each paired joint $j\in\mathcal{J}_{dual}$:
\begin{equation}
    \mathcal{L}_{sym}
    =
    \frac{1}{L|\mathcal{J}_{dual}|}\sum_{n=1}^{L}\sum_{j\in\mathcal{J}_{dual}}
    \left(\hat{d}_{n,j}-d_{n,j}\right)^2,
\end{equation}
where $d_{n,j}=\left\|\mathbf{y}^{L}_{n,j}-\mathbf{y}^{R}_{n,j}\right\|_2$ is the inter-hand distance. We set $\lambda_{geo} = 0.1$ and $\lambda_{sym} = 0.2$.

\paragraph{Classifier-Free Guidance.}
During training, we randomly discard the conditioning signals (both the body context $\mathbf{b}_{seq}$ and history buffer $\mathbf{h}_{past}$) with probability $p{=}0.1$, replacing them with a learnable null embedding $\emptyset$. This masking strategy serves a dual purpose: (1) it allows the model to learn the intrinsic hand motion manifold independently of external signals; and (2) masking $\mathbf{h}_{past}$ simulates the cold-start scenario, enabling autoregressive rollout to initiate purely from the body context without requiring a ground-truth history buffer. At inference, we extrapolate the model prediction using a guidance scale $s > 1.0$:
\begin{equation}
    \hat{\mathbf{x}}_{pred} = \hat{\mathbf{x}}_\theta(\mathbf{z}_t, \emptyset) + s \cdot \left( \hat{\mathbf{x}}_\theta(\mathbf{z}_t, \mathbf{b}_{seq}, \mathbf{h}_{past}) - \hat{\mathbf{x}}_\theta(\mathbf{z}_t, \emptyset) \right).
\end{equation}
A scale $s > 1$ effectively sharpens the conditional distribution, ensuring the generated hands are tightly coupled with the body dynamics and temporally continuous with the history.
\rev{This single masking mechanism plays a dual role: at training time it teaches both the unconditional manifold and the cold-start behavior (when no past hand motion is available), and at inference time the same conditional/unconditional pair enables classifier-free guidance, trading off body consistency against semantic responsiveness via $s$ without retraining.}

\paragraph{Implementation Details.}
We use a DDPM sampler with direct $\hat{\mathbf{x}}_0$ prediction and a 50-step cosine noise schedule. The history window is set to $P{=}10$, the future prediction window to $L{=}45$, and the latent dimension to 256. This configuration achieves real-time performance (450+ FPS) on a single RTX 4090.

\subsection{Semantically-Layered Conditional Adaptation}
\label{subsec:adaptation}

We implement manifold restriction by freezing the kinematic prior and learning only lightweight adapters/gates for conditioning, so downstream control steers trajectories within the learned kinematic distribution rather than constructing motion from sparse supervision.
Once the prior $\epsilon_\theta$ is trained, it captures the manifold of natural hand-body dynamics. For downstream tasks (e.g., text-driven generation), we aim to steer this manifold without breaking its kinematic consistency.

Standard adaptation methods often inject conditioning signals uniformly across all network layers. However, having established a kinematic hierarchical structure in our prior, we observe that semantic commands also operate at different kinematic levels. For instance, instructions like ``energetically'' imply a modulation of momentum transfer (affecting upstream spine/shoulder layers), while ``pinch with two fingers'' targets local articulation (affecting downstream hand layers).

To exploit this structural alignment, we propose a semantically-layered adapter. Instead of applying a monolithic control signal, we introduce layer-specific learnable gates $\{\gamma_l\}_{l=1}^L$ that modulate the injection of the condition embedding $\mathbf{c}$ (e.g., text or attributes) into the $l$-th cascading stage of the frozen prior:
\begin{equation}
    \hat{\epsilon}(\mathbf{x}_n^{(\tau)}, \dots)_l = \epsilon_\theta(\mathbf{x}_n^{(\tau)}, \dots)_l + \text{Tanh}(\gamma_l) \cdot \phi_l(\mathbf{x}_n^{(\tau)}, \mathbf{c}),
\end{equation}
where $\gamma_l$ is initialized to zero. This design allows the network to automatically disentangle high-level dynamic styles from fine-grained pose constraints. 

In practice, we provide two lightweight control instantiations using the same adapter formulation: (i) \emph{attribute-driven control}, where hand-local attributes (e.g., finger spread or openness) are injected only into late stages to preserve upstream body--hand consistency while achieving precise local modulation; and (ii) \emph{text-driven control}, where the adapter can be enabled across a broader set of stages to affect both global dynamics and fine-grained articulation. Crucially, being able to dynamically select which layers receive the condition makes adaptation substantially more data-efficient: the model only learns the minimal residual needed for the target control, instead of re-learning kinematics end-to-end under scarce labels.

\subsection{Conditional Adaptation Strategies}
\label{subsec:adaptation_strategies}

Having defined the adapter architecture, we now focus on its training. 
Since the heavy lifting of kinematic modeling is handled by the frozen prior, the adaptation phase does not require learning physics from scratch. This allows us to adopt a highly data-efficient training protocol.
In this section, we first outline the governing principles of this efficiency, followed by two specific instantiations: self-supervised attribute control and weak-supervised text control.

\paragraph{Data Efficiency and Sampling}
\label{subsec:data_principles}

\begin{figure}[t]
	\centering
	\includegraphics[width=\linewidth]{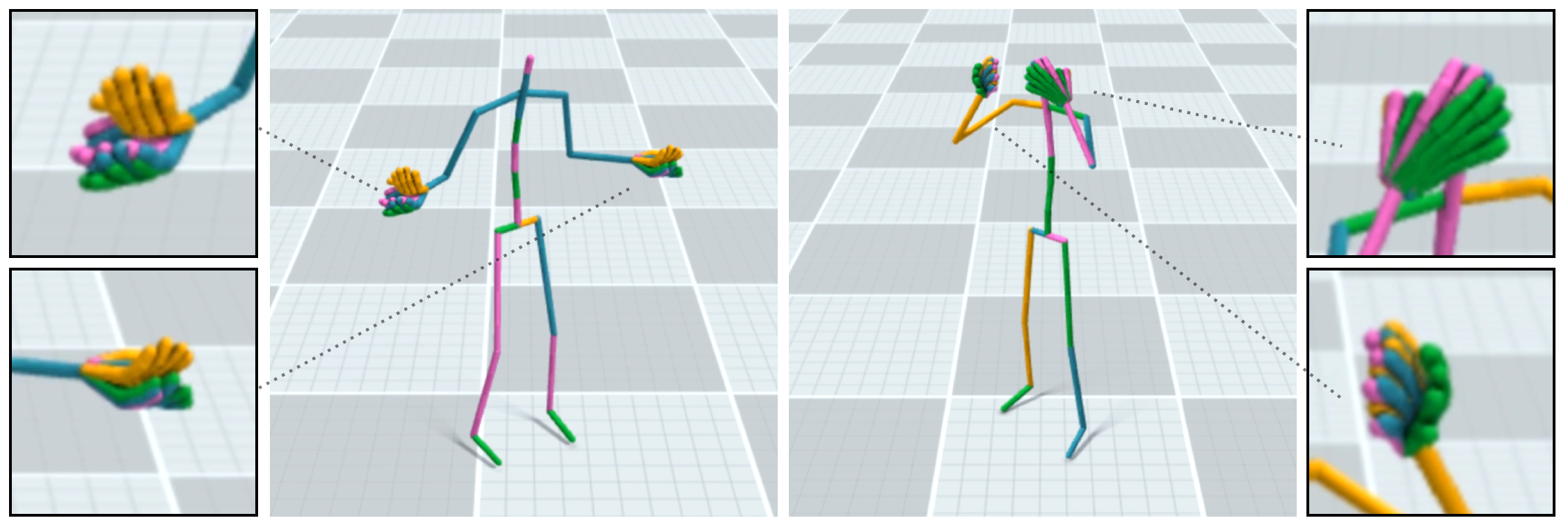}
	\caption{
        Same body motion input can generate distinct yet plausible hand motion by given different random seeds. 
        }
    \label{fig:sampling_diversity}
\end{figure}

A central premise of our work is that the kinematic prior already captures the physics of motion (how to move). As shown in Figure~\ref{fig:sampling_diversity}, our prior models a distribution of plausible hand completions rather than overfitting to a single memorized clip, leaving the adapter to learn only the intent (why to move). 
Empirically, we observe an order-of-magnitude reduction in the data scale required for adaptation. While our kinematic prior is trained on $\sim$100 hours of motion to cover the kinematic manifold, conditional adapters converge with as little as a few hours of labeled data.
This confirms that the adapter's role is strictly limited to manifold navigation (selecting valid trajectories) rather than manifold construction. 

In the conditional stage, our adapters steer this diversity toward the user-specified control signal by aligning the generated sample with the intended attribute or text condition while preserving body consistency.

\paragraph{Sampling Strategy.}
To maximize manifold coverage, we employ a clustering-based sampling strategy.
We extract the input body motion features from the available pool and perform K-Means clustering ($K=50$). We then sample training instances uniformly across clusters.
This ensures that the sparse injected training set spans the full range of body dynamics, forcing the adapter to learn a generalized mapping valid across the prior's entire latent space.

\subsubsection{Application I: Attribute-Driven Control}
\label{subsec:attr_control}

We define scalar attribute functions to map the ground-truth hand motion $\mathbf{x}$ to a descriptor $\mathbf{a}$ directly from the dataset. Since these attributes are derived algorithmically, we obtain training pairs $(\mathbf{b}, \mathbf{x}, \mathbf{a})$ automatically via self-supervision without manual annotation.

\paragraph{Example: fist tightness (clench level).}
Using FK, we convert rotations to global joint positions $\mathbf{y}_{n}=\mathrm{FK}(\mathbf{b}_{n},\mathbf{x}_{n};\mathcal{S})$.
For each finger $f\in\{1,\dots,5\}$, we measure the distance between its fingertip and proximal joint:
\begin{equation}
    d_{n,f}=\left\|\mathbf{y}_{n,\mathrm{tip}(f)}-\mathbf{y}_{n,\mathrm{base}(f)}\right\|_2 .
\end{equation}
A more clenched finger yields a smaller $d_{n,f}$. We aggregate a clip-level tightness score by averaging across fingers and time:
\begin{equation}
    a = \frac{1}{L}\sum_{n=1}^{L}\left(\frac{1}{5}\sum_{f=1}^{5} d_{n,f}\right),
\end{equation}
then discretize $a$ into $K$ levels and encode it as a one-hot vector $\mathbf{a}\in\{0,1\}^{K}$.

Given that these attributes are spatially localized to the hand, we constrain injection exclusively to the last two layers of the prior, which correspond to the fine-grained hand and finger kinematic levels. By masking injection for all upstream layers, the global body consistency established by the prior remains strictly preserved, while allowing precise modulation of hand articulation.
\rev{We note that the attribute condition is currently applied at the clip/window level: a single scalar level is held fixed over the controlled segment rather than specified per frame.}

\subsubsection{Application II: Text-Driven Control}
\label{subsec:text_control}

For intuitive intent control, we utilize text prompts. Unlike single-scalar attributes, text descriptions often encompass both global dynamics (e.g., ``waving'') and local articulation (e.g., ``fingers splayed''). To address the scarcity of hand-text datasets, we render body-hand motions into videos and query a Vision-Language Model (QWEN-VL) for captioning. Given a closed-set action vocabulary, we enforce a structured prompting template to query for the global action category, hand-object interaction type, and fine-grained finger pose.

\paragraph{Hand verb vocabulary.}
We define a compact verb lexicon for captioning and prompting:
(i) \emph{deictic}: point, indicate, gesture, direct, holding;
(ii) \emph{social}: wave, greet, beckon, invite, dismiss, salute;
(iii) \emph{iconic/emphatic}: emphasize, present, show, demonstrate;
(iv) \emph{counting/symbolic}: count, enumerate, signal, pose;
(v) \emph{self-contact}: touch, rub, tap, scratch, clasp;
(vi) \emph{two-hand}: clap, hold-hands, interlace;
(vii) \emph{object-like}: grab, pinch, grasp, release, open, close, squeeze.
Hand state modifiers include: spread, splayed, flat, relaxed, stiff, clenched, curled, cupped, neutral.

\paragraph{Full-layered Semantic Injection.}
In contrast to the attribute-driven strategy (Sec.~\ref{subsec:attr_control}) where injection is constrained to the final layers, here we enable the adapter across the full kinematic hierarchy.
We inject the text embedding $\mathbf{c}_{text}$ into all layers of the prior, initializing all gates $\gamma_l=0$ and allowing them to be optimized end-to-end.
Empirically, we observe that the gates $\gamma_l$ in upstream layers tend to activate for global action verbs (modulating momentum), while downstream gates activate for specific noun descriptors (modulating finger articulation). This validates the effectiveness of our layered design in disentangling complex textual intents.

\begin{table*}[t]
    \centering
    \caption{Body-hand prior evaluation (Dataset A test split). We report statistical metrics (FID, Diversity, FPS), physical \& geometric consistency (MPJPE, Root Error, Velocity Smoothness/jerk), and perceptual scores (Naturalness, Dynamic Richness; 1--5). Arrows indicate the preferred direction.}
    \label{tab:prior_eval}
    \resizebox{0.8\textwidth}{!}{
    \begin{tabular}{l|cccccc|cc}
        \toprule
        Method & FID $\downarrow$ & Diversity $\uparrow$ & FPS $\uparrow$ & MPJPE (cm) $\downarrow$ & Root Err(deg) $\downarrow$ & Vel.\ Smooth. $\downarrow$ & Naturalness (1--5) $\uparrow$ & Dyn.\ Richness (1--5)   $\uparrow$ \\
        \midrule
        MDM~\citeyearpar{mdm} &   6.593   &  1.195   &   30   &   1.807   &   7.065  &   0.293   &   3.68   &   3.10   \\
        PriorMDM~\citeyearpar{shafir2024human}  &   22.646   &   0.672   &   30   &   2.695   &   20.177   &   0.524   &   1.93   &   2.08   \\
        Body2Hands~\citeyearpar{ng2021body2hands} &   12.601   &   0.342   &   65   &   2.432   &   11.988   &   0.173   &  2.95    &   2.46   \\
        BOTH2Hands~\citeyearpar{zhang2024both2hands} &   6.179   &   1.051   &   12 (1000-step)   &   1.926   &   6.742   &   0.118   &   3.04   &   3.21   \\
        \midrule
        Our AR-based Prior &   \textbf{3.691}   &   \textbf{1.742}   &   450   &   \textbf{1.310}   &   \textbf{5.436}   &   \textbf{0.035}   &   \textbf{4.18}   &   \textbf{4.38}   \\
        \bottomrule
    \end{tabular}
    }
\end{table*}

\section{Evaluation}
\label{sec:eval}

Our framework is built on a central hypothesis: decoupling kinematic learning from semantic conditioning yields better hand motion quality, stronger generalization, and higher data efficiency than end-to-end conditioned generation. We design our evaluation to test this hypothesis from multiple angles: (1) whether the unsupervised prior alone produces kinematically plausible hand motion that outperforms end-to-end baselines (Sec.~\ref{subsec:prior_eval}); (2) whether lightweight adapters can achieve competitive controllability with limited paired supervision (Sec.~\ref{subsec:cond_eval}); (3) whether the prior transfers across datasets better than models trained from scratch (Sec.~\ref{subsec:cross_dataset}); and (4) whether each architectural component contributes meaningfully (Sec.~\ref{subsec:ablation}).

\subsection{Setup}

\paragraph{Datasets.}

We construct three distinct datasets to rigorously test different aspects of our model:
\begin{itemize}
    \item Dataset A: From multiple public sources (MotionPersona, InterAct, BEAT) and in-house captures, we collect and open-source a large-scale motion capture dataset containing 100 hours of diverse motions, covering locomotion, daily activities, and dyadic interactions. We additionally apply left--right mirroring to double the effective training set to 200 hours. Detailed composition is provided in the supplementary material. This dataset is shuffled and split into training and testing sets at the clip level with an 8:2 ratio. It is used exclusively to train the frozen prior $\epsilon_\theta$. 
    \item Dataset B: We labeled a 3-hour subset of the training data with text descriptions, and additional 30 minutes as test set. This subset serves as the scarce paired data for training the conditional adapters. 
    \item Dataset C: To evaluate robustness on in-the-wild motions, we curated 1,000 clips from AMASS (diverse sampling) and HunyuanMotion (text-to-motion outputs). These samples lack ground-truth hand motions and are used for qualitative evaluation and user scoring.
\end{itemize}

\begin{figure}[h]
    \centering
    \includegraphics[width=\linewidth]{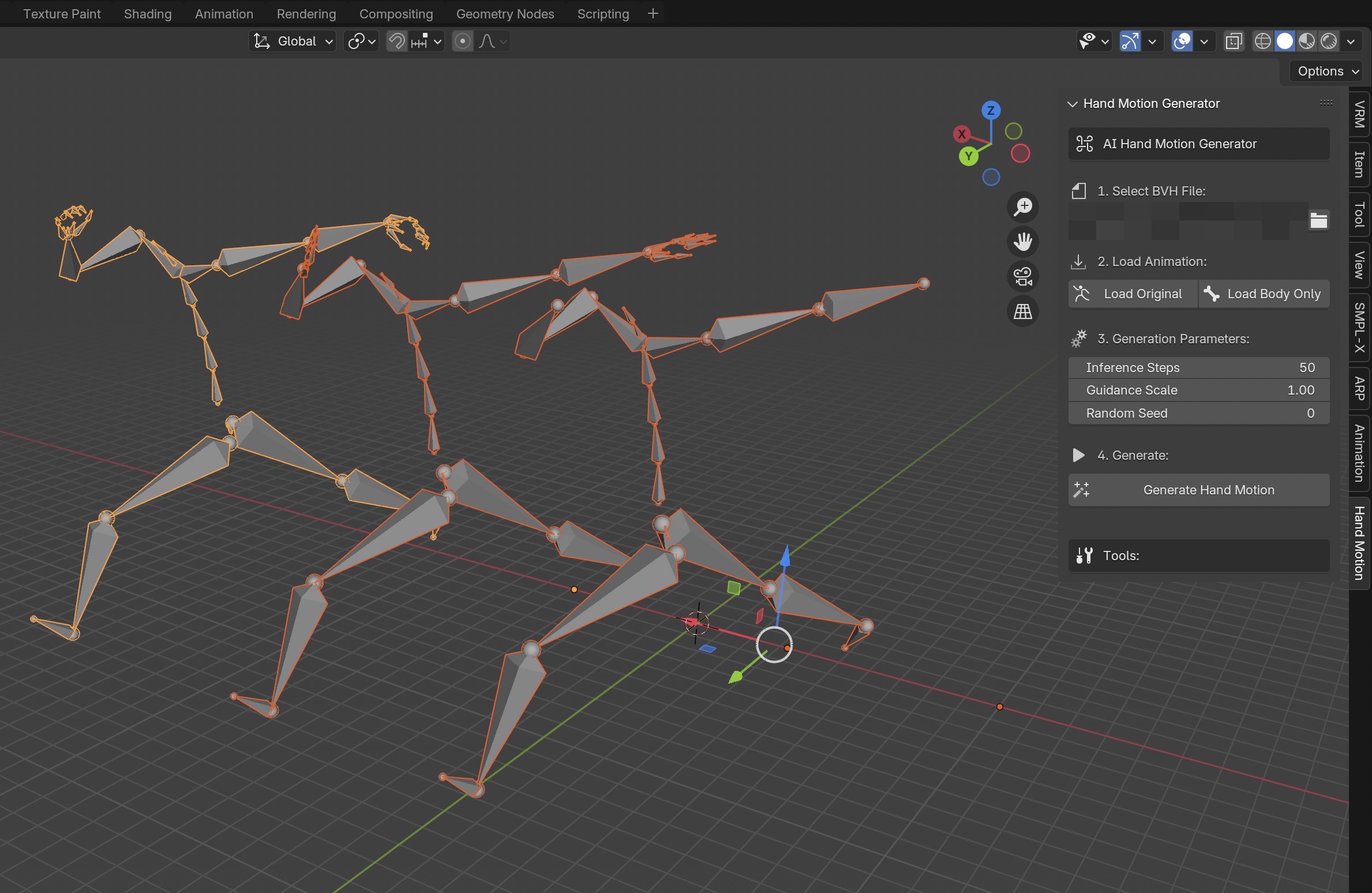}
    \caption{Runtime screenshot of the Blender add-on. Left to right: our generation, identity hands, original input.}
    \label{fig:blender_addon_main}
\end{figure}

Note that since our model operates on a specific kinematic tree structure, all input BVH files are retargeted to our unified skeleton template. We resample all motion data to 30 FPS. For the in-the-wild dataset, we perform root-relative centering before inference. We also provide a Blender add-on for interactive authoring (Figure~\ref{fig:blender_addon_main}), which supports efficient generation (9000 frames in 15 seconds on a MacBook CPU).

\paragraph{Metrics.}
We propose a comprehensive evaluation protocol covering three dimensions: (1) Statistical metrics: \textbf{FID} (Fr\'{e}chet Inception Distance, computed in a learned hand-motion embedding space via a pretrained autoencoder) to measure the distributional distance to the ground truth, \textbf{Diversity} to measure the variance of generated samples (a score ${\geq}0.7$ typically corresponds to qualitatively distinct hand configurations such as fist vs.\ splayed fingers, while ${\leq}0.3$ indicates near-identical samples), and \textbf{FPS} to measure inference speed on a single RTX 4090. (2) Kinematic \& Geometric Consistency: We compute \textbf{MPJPE} (Mean Per-Joint Position Error) for hand end-effectors. Crucially, to validate our kinematic chain modeling, we report \textbf{Root Error} (orientation error of the wrist relative to the forearm) and \textbf{Velocity Smoothness} (mean jerk). A lower Root Error explicitly indicates better phase-locking with the parent kinematic chain. (3) Perceptual scoring: We deploy a batch scoring protocol where human evaluators rate samples on a Likert scale (1--5). We report \textbf{Naturalness} (kinematic plausibility), \textbf{Dynamic Richness} (avoidance of mean-pose collapse), and for conditional tasks, \textbf{Text--Motion Consistency}.

\subsection{Evaluation on Prior Learning}
\label{subsec:prior_eval}

The first and most fundamental question is whether a body-conditioned prior, trained without any semantic supervision, can already produce high-quality hand motion. If the prior captures the intrinsic body--hand kinematic distribution well, it should outperform methods that jointly learn kinematics and semantics from limited data. We evaluate this by comparing body-conditioned hand completion on the held-out test split of Dataset A, where all methods receive only the body motion as input.

\paragraph{Baselines.}
We compare against four learning-based methods spanning different paradigms: (i) MDM~\cite{mdm}, a diffusion-based full-body model adapted to use body motion as the conditioning signal (50 denoising steps, 150 fixed frames); (ii) PriorMDM~\cite{shafir2024human}, which reconstructs hand motion by masking clean body parts during denoising (150 fixed frames); (iii) Body2Hands~\cite{ng2021body2hands}, a deterministic regression model that directly maps upper-body motion to hand pose; and (iv) BOTH2Hands~\cite{zhang2024both2hands}, a diffusion model that generates hand motion conditioned on body dynamics. For fair comparison, we strip all additional conditioning signals (text, audio, etc.) from the baselines, training them purely to map body dynamics to hand motion.

We also implemented a nearest-neighbor (NN) motion-graph baseline following J\"{o}rg et al.~\cite{jorg2012data}, using the same body features as our method. While NN retrieval can be competitive in small, constrained domains (e.g., a few hours of co-speech gestures), it cannot simultaneously achieve scalability, real-time runtime, and high quality under large, dynamic motions. Concretely, scaling the database is memory-intensive (even with 128\,GB RAM we can only load ${\sim}20$ hours for accelerated matching); enforcing temporal continuity via graph optimization is slow ($>$3 minutes for 9000 frames with 5 hours of data); and when large body dynamics are involved, stitching inevitably introduces visible artifacts at segment boundaries. For these reasons, we include NN as a non-generative reference point but focus the quantitative comparison on learning-based methods.

\paragraph{Quantitative Results.}
Table~\ref{tab:prior_eval} shows our prior is best across all three dimensions: distributional quality (lowest FID, highest Diversity), geometric consistency (lowest MPJPE, Root Error, jerk), and perceptual quality (highest Naturalness and Dynamic Richness). Root Error and jerk are especially relevant to KCCA: lower Root Error reflects better wrist--forearm coupling, and low jerk confirms the autoregressive rollout introduces no discontinuities at window boundaries.

\begin{figure*}[t]
	\centering
	\includegraphics[width=\linewidth]{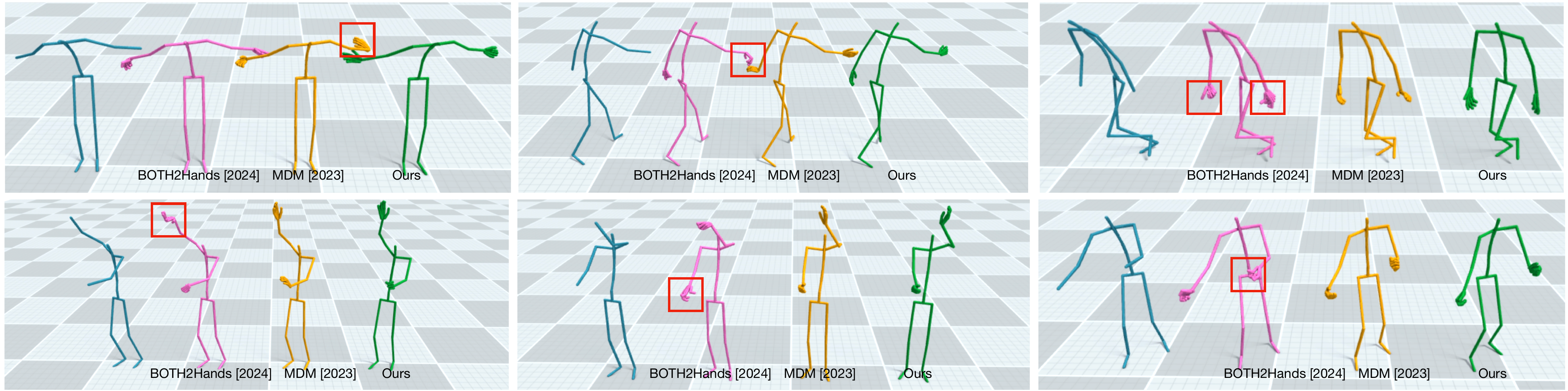}
	\caption{Qualitative comparison of our prior with baselines on the test split of Dataset A. Our method produces stable wrist and plausible finger articulation, whereas BOTH2Hands exhibits unstable wrist behavior and MDM occasionally generates unnatural joint rotations.}
    \label{fig:comparison_prior}
\end{figure*}

Figure~\ref{fig:comparison_prior} shows the qualitative gap. BOTH2Hands often exhibits unstable wrist attachment or implausible finger configurations, likely because its training distribution lacks large-range body motions, and MDM produces unnatural rotations once extended to hands -- naively adding high-DoF hand generation degrades end-to-end diffusion baselines. Deterministic regressors like Body2Hands average variation away into a single ``safe'' pose (mean-pose collapse), whereas our probabilistic model keeps diverse, plausible completions. Only our model and Body2Hands reach real-time speed (60+ FPS).

\begin{table*}[t]
    \centering
    \caption{Text-driven control on the test split of Dataset B (30 minutes). All methods are trained on the same 100-hour motion pool; for clips without text labels, end-to-end baselines use a null (empty) text condition during training.}
    \label{tab:cond_eval}
    \resizebox{0.9\textwidth}{!}{
    \begin{tabular}{l|ccccc|ccc}
        \toprule
        Method & FID $\downarrow$ & Diversity $\uparrow$ & MPJPE (cm) $\downarrow$ & Root Err(deg) $\downarrow$ & Vel.\ Smooth. $\downarrow$ & Naturalness (1--5) $\uparrow$ & Dyn.\ Richness (1--5)   $\uparrow$ & Text--Motion Consistency (1--5) $\uparrow$ \\
        \midrule
        MDM-Fullbody~\citeyearpar{mdm} &  5.161   &   0.633   &  1.095   &   6.904  &   0.169  &   3.07  &   2.55  &   3.04  \\
        BOTH2Hands~\citeyearpar{zhang2024both2hands} &   4.975   &   0.659   &   1.257   &   6.879   &    0.148  &   3.16  &    2.71   &   3.21   \\
        \midrule
        Ours (end-to-end) &  3.026  &   \textbf{0.971} &   0.950  &   5.342 &  0.106 &   3.76  &   3.31 &    3.56 \\
        Ours (Adapter) &  \textbf{2.531}  &  0.897  &   \textbf{0.863}   &  \textbf{4.021}  &   \textbf{0.087}  &   \textbf{3.98}  &   \textbf{3.54}  &   \textbf{4.21}  \\
        \bottomrule
    \end{tabular}
    }
\end{table*}

\subsection{Evaluation on Conditional Generator Learning}
\label{subsec:cond_eval}

Having shown the prior produces high-quality unconditional completions, we now ask whether lightweight adapters add semantic control without degrading kinematic quality. This tests the \emph{prior-first, condition-second} hypothesis: if the adapter only learns ``where on the manifold to go'' rather than reconstructing kinematics, it should match end-to-end controllability with far less labeled data.

\paragraph{Baselines.}
We compare against end-to-end MDM-Fullbody and BOTH2Hands. All share the same 100-hour pool, with only a small text-labeled subset (Dataset B) and the rest unlabeled; end-to-end baselines use the text condition when available and a null condition otherwise. To isolate the adapter, we include ``Ours (end-to-end),'' which trains our model end-to-end instead of freezing the prior. Evaluation uses the held-out labeled split of Dataset B.

\begin{figure}[t]
    \centering
    \includegraphics[width=\linewidth]{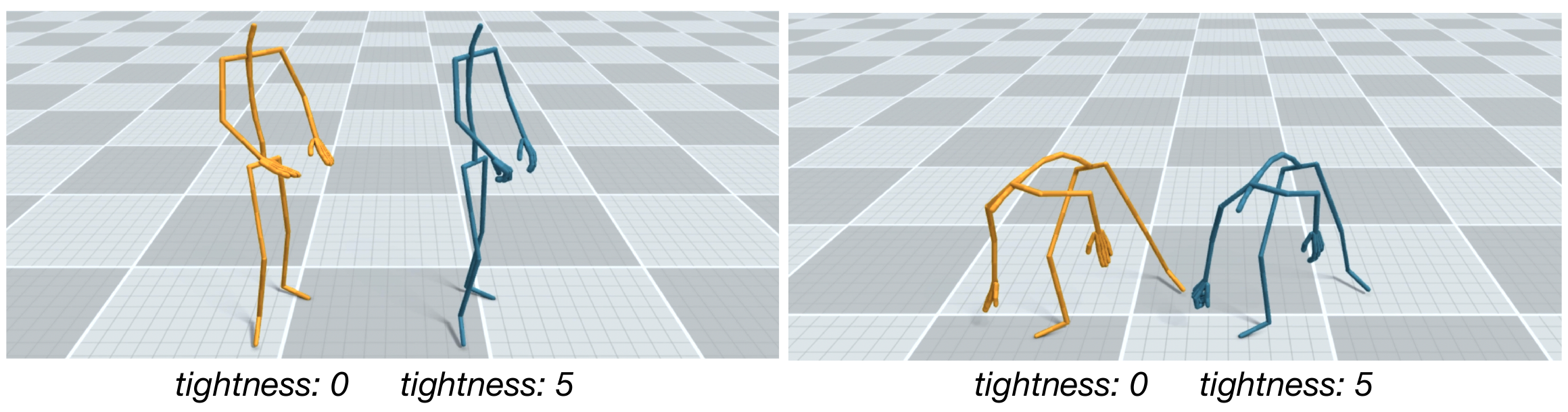}
    \caption{Attribute-driven control: varying the fist tightness level while keeping body motion fixed.}
    \label{fig:cond_attribute}
\end{figure}

\begin{figure}[t]
    \centering
    \includegraphics[width=\linewidth]{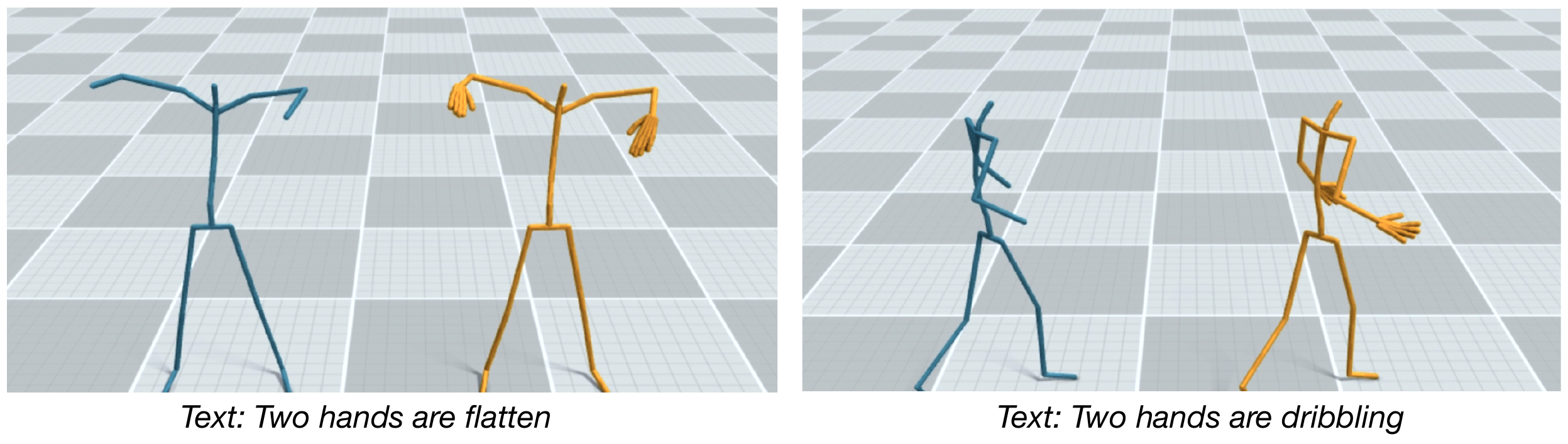}
    \caption{Text-driven control: a text prompt steers both global gesture dynamics and fine-grained finger articulation.}
    \label{fig:cond_text}
\end{figure}

\paragraph{Quantitative Results.}
Table~\ref{tab:cond_eval} shows our adapter achieves the best trade-off across distributional quality, kinematic accuracy, and controllability. The comparison with ``Ours (end-to-end)'' is telling: both share the same architecture, but the adapter builds on a frozen prior that anchors generation to the kinematic manifold, whereas end-to-end training must jointly fit unconditional quality and conditional control from mixed supervision -- under scarce labels this often leads to under-using the text condition. The external baselines lag further, reflecting the difficulty of learning kinematics and semantics jointly from limited paired data. More results are in Figures~\ref{fig:cond_attribute} and~\ref{fig:cond_text}.

\begin{figure}[h]
    \centering
    \includegraphics[width=\linewidth]{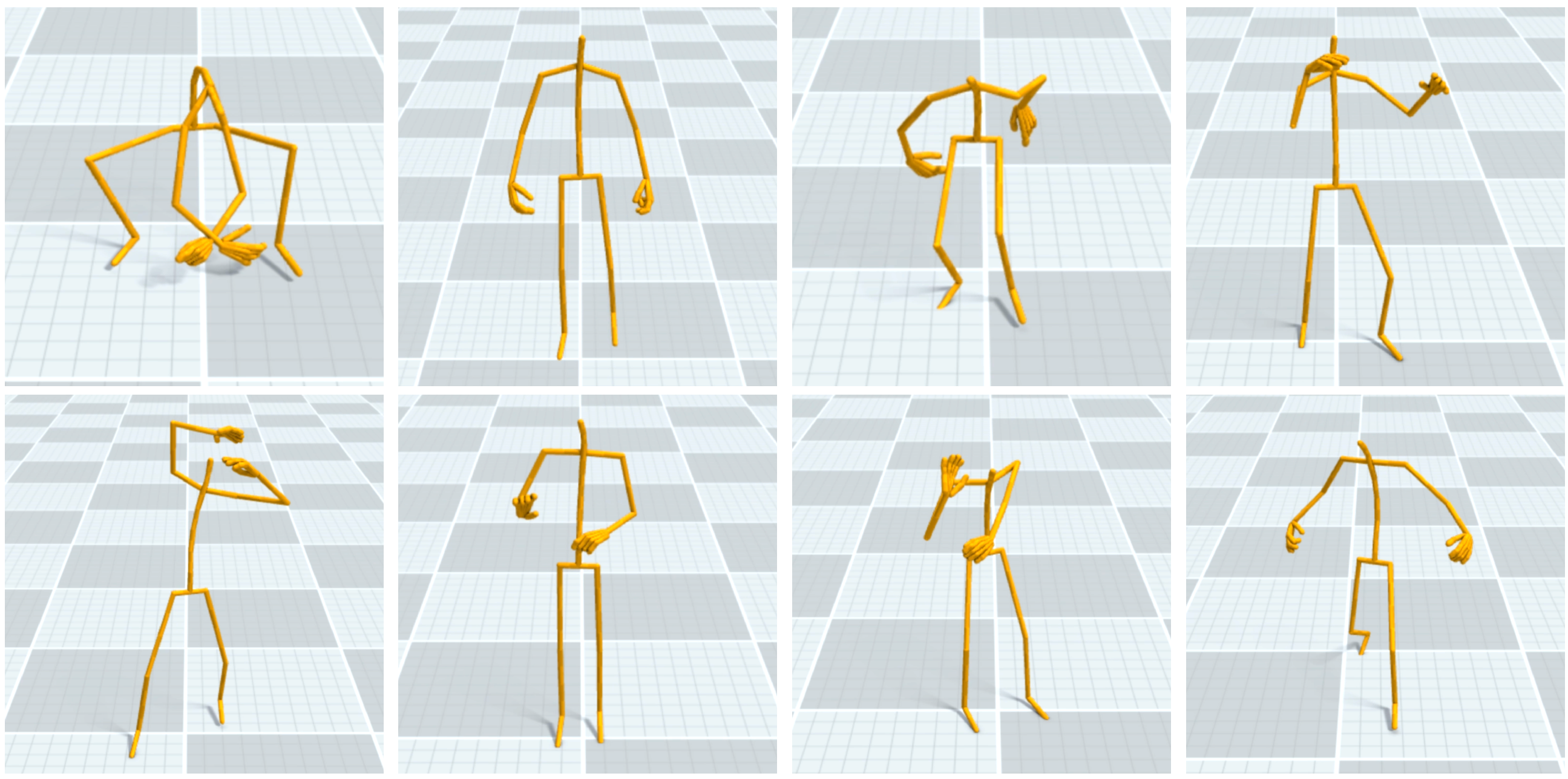}
    \caption{Robustness on in-the-wild body motions from HunyuanMotion~\cite{hymotion2025}.}
    \label{fig:wild_results}
\end{figure}

\paragraph{Generalization to In-the-Wild Motions.}
A practical system must handle out-of-distribution body motions. We test on Dataset C (AMASS and HunyuanMotion~\cite{hymotion2025} text-to-motion outputs), unseen during training and without ground-truth hands. As shown in Figure~\ref{fig:wild_results}, our prior still produces plausible hands consistent with the upstream arm dynamics, indicating that the learned manifold generalizes beyond the training distribution and can be paired with external body-motion generators without hand-specific supervision.

\subsection{Cross-Dataset Generalization}
\label{subsec:cross_dataset}

The previous experiments use our own Dataset A for prior training and Dataset B for adaptation, which may raise the question of whether the improvements are specific to our data pipeline. To directly test cross-dataset generalization, we design a controlled transfer experiment: we take the BOTH2Hands dataset (2.09\,h labeled training data, 0.22\,h fixed test split) as an entirely separate target domain (Dataset B$'$) and evaluate how well a prior trained on our Dataset A transfers to this new domain compared with models trained from scratch on B$'$ alone.

We systematically vary the labeled data budget $K\in\{0.25\text{h},\;1.0\text{h},\;2.09\text{h}\}$ to simulate increasingly data-scarce scenarios, and compare four configurations: (i) Ours $A{\to}B'$ (prior trained on ${>}100$\,h Dataset A, adapted on B$'$); (ii) Ours B$'$-only (prior and adapter trained solely on B$'$); (iii) BOTH2Hands and (iv) MDM, both trained end-to-end on B$'$.

\begin{table}[b]
    \centering
    \caption{Cross-dataset generalization on the BOTH2Hands test split. Columns correspond to different labeled data budgets $K$.}
    \label{tab:cross_dataset}
    \resizebox{\linewidth}{!}{%
    \begin{tabular}{l|ccc|ccc|ccc}
        \toprule
        & \multicolumn{3}{c|}{FID $\downarrow$} & \multicolumn{3}{c|}{Diversity $\uparrow$} & \multicolumn{3}{c}{MPJPE (cm) $\downarrow$} \\
        Method & 2.09h & 1.0h & 0.25h & 2.09h & 1.0h & 0.25h & 2.09h & 1.0h & 0.25h \\
        \midrule
        MDM          & 5.83 & 6.83 & 8.01 & 0.362 & 0.227 & 0.148 & 1.24 & 1.52 & 2.03 \\
        BOTH2Hands   & 4.93 & 5.96 & 7.95 & 0.390 & 0.293 & 0.171 & 1.18 & 1.44 & 1.91 \\
        Ours B$'$-only & 4.05 & 4.87 & 7.60 & 0.405 & 0.316 & 0.202 & 1.12 & 1.36 & 1.82 \\
        Ours $A{\to}B'$ & \textbf{2.59} & \textbf{2.97} & \textbf{3.85} & \textbf{0.938} & \textbf{0.906} & \textbf{0.810} & \textbf{0.88} & \textbf{0.93} & \textbf{1.01} \\
        \bottomrule
    \end{tabular}
    }
\end{table}

As shown in Table~\ref{tab:cross_dataset}, end-to-end baselines degrade sharply as $K$ decreases (FID nearly doubles, MPJPE rises substantially at $K{=}0.25$\,h), indicating overfitting to the shrinking set, whereas Ours $A{\to}B'$ degrades gracefully and is best at every budget. The Diversity gap is striking: at $K{=}0.25$\,h end-to-end baselines collapse to near-zero diversity while our transferred prior keeps strong variety, confirming the learned manifold resists mode collapse under minimal adaptation data.
Notably, even Ours B$'$-only outperforms both end-to-end baselines, highlighting that the prior--adapter factorization itself is beneficial regardless of pretraining scale.

\subsection{Ablation Studies}
\label{subsec:ablation}

We now isolate the contribution of each design choice to understand what drives the improvements observed above.

\begin{figure}[h]
    \centering
    \includegraphics[width=\linewidth]{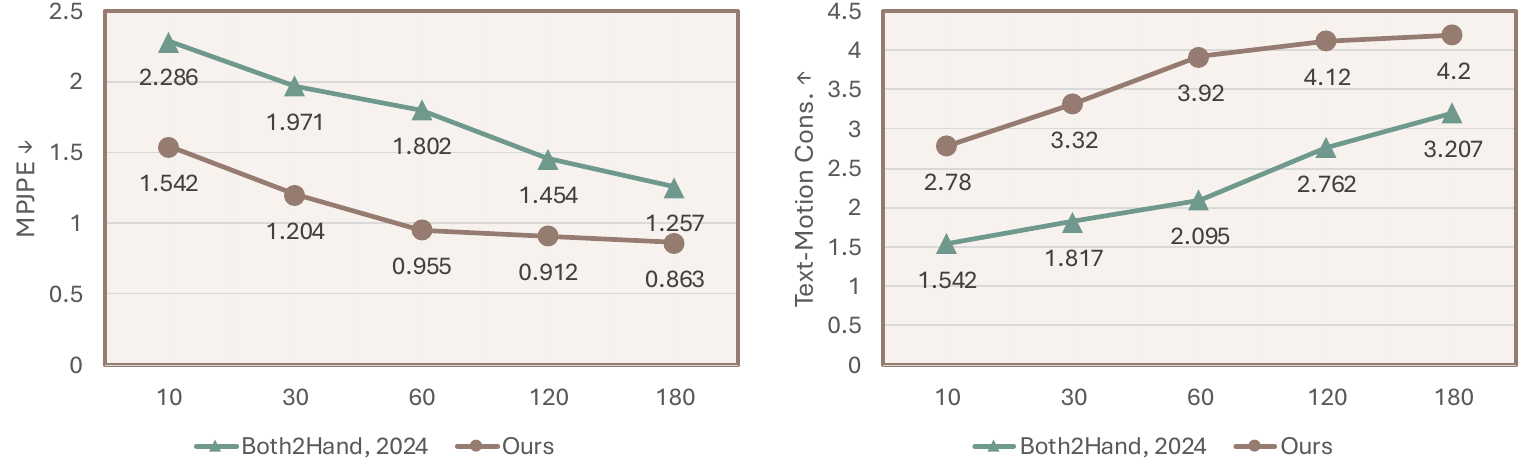}
    \caption{Data Efficiency for Adapter. We compare different scales of labeled data from Dataset B \rev{(horizontal axis in minutes)}, and report MPJPE and Text--Motion Consistency after the injection.}
    \label{fig:ablation_data}
\end{figure}

\paragraph{Data Efficiency for Adapter.}
Since the adapter learns manifold navigation rather than kinematic reconstruction, it should need far less labeled data. We train it on subsets of Dataset B (10 minutes to 3 hours) and compare against BOTH2Hands. As shown in Figure~\ref{fig:ablation_data}, \rev{our adapter's accuracy and controllability largely saturate after only a few hours of labeled data, indicating that little supervision is needed} when the kinematic foundation is in place, whereas BOTH2Hands keeps improving gradually, reflecting the burden of jointly learning kinematics and semantics.

\paragraph{Module Effectiveness.}
We ablate KCCA and semantically-layered injection (Table~\ref{tab:ablation_modules}). KCCA mainly improves kinematic accuracy (MPJPE, Root Error) by strengthening wrist--forearm coupling but adds little controllability on its own; layered injection is the key driver of Text--Motion Consistency, placing conditions at the right kinematic levels so the adapter learns a more targeted residual. The full model is best overall. Removing the dual-hand distance loss slightly improves Root Error but increases MPJPE, indicating it usefully regularizes hand position.

\begin{table}[h]
    \centering
    \caption{Ablation on attention and injection strategy. We report hand reconstruction error (MPJPE), wrist--forearm alignment (Root Error), and text--motion consistency (1--5) after the conditional injection.}
    \resizebox{0.95\linewidth}{!}{%
    \begin{tabular}{lccc}
        \toprule
        Setting & MPJPE $\downarrow$ & Root Err $\downarrow$ & Text--Motion Consistency (1--5) \\
        \midrule
        w/o KCCA, w/o Layered &  0.950    &   5.342   &   3.25   \\
        w/ KCCA, w/o Layered &  0.867    &   4.217   &   3.45   \\
        w/o reflection loss & 0.879 & \textbf{4.011} & \textbf{4.21} \\
        \midrule
        Ours (KCCA + Layered) &  \textbf{0.863}    &   4.021   &   \textbf{4.21}   \\
        \bottomrule
    \end{tabular}
    }

    \label{tab:ablation_modules}
\end{table}

\paragraph{Lower-Body Features.}
One might expect that only upper-body joints matter for hand prediction. However, removing lower-body features from the body conditioning degrades all metrics, confirming that global momentum cues from locomotion and turns propagate to upper-limb dynamics and benefit hand completion.

\section{Discussion \& Limitations}
\label{sec:discussion}

The main contribution of this work is the practical validation of the \emph{prior-first, condition-second} principle for body-conditioned hand motion completion. Since semantic supervision for motion is scarce, expensive, and dataset-specific, learning a kinematic prior from abundant motion and adding conditioning via lightweight adaptation is an effective strategy under limited supervision.

\paragraph{Limitations and Future Work.}
A key limitation is scaling to full-body generation with the same representation. Hand motion is largely governed by local rotations and lies in a compact manifold that suits our rotation-only prior; full-body motion involves global transport and long-horizon coupling that this representation captures less well, suggesting the need for task-appropriate priors rather than one model for all subproblems. When conditioning conflicts with body dynamics (e.g., ``jazz hands'' during vigorous dancing), the classifier-free guidance scale balances semantics against body conditioning, and the strong prior keeps results stable with fewer artifacts than end-to-end baselines. Our framework also does not explicitly model hand--object interaction or contact; adding object state and contact objectives could be an important future work.

\rev{Our attribute-based control is also coarse in time: a single control level is held fixed over the clip than specified frame. While time-varying control is in principle achievable by changing the level across autoregressive windows, our current system does not expose a polished per-frame or trajectory-based control interface. }

\rev{A further limitation is semantic, rather than kinematic, correctness: our model optimizes for kinematic plausibility and body--hand consistency but does not reason about object state or task success. Hence in object-centric actions (e.g., a basketball shot, where the hand should open at release) the motion can be kinematically plausible yet semantically wrong, an error our distributional and geometric metrics do not capture. Action-correctness or interaction-aware evaluation, together with object/contact conditioning, is thus an important complement to our setting.}

\section{Conclusion}
\label{sec:conclusion}

We introduced a two-stage framework for controllable hand motion completion conditioned on body motion, decoupling a learned autoregressive body--hand kinematic prior from a semantically-layered adapter that enables diverse control with limited paired supervision. Together they provide a practical path to real-time, controllable hand completion in animation pipelines while keeping kinematic plausibility grounded in the learned prior.

\section*{Acknowledgements}
This work was partially funded by the Innovation and Technology Commission of the HKSAR Government under the ITSP-Platform grants (Ref: ITS/335/23FP, ITS/469/24FP).

\bibliographystyle{eg-alpha-doi}
\bibliography{main}

\end{document}